\documentclass{article}
\usepackage[T1]{fontenc} 
\usepackage{amsmath,amscd,amsfonts,amssymb,amsthm,stmaryrd}
\usepackage{latexsym}
\usepackage{verbatim}
\usepackage{graphicx}
\usepackage{epsfig}
\usepackage[english]{babel}
\usepackage{euscript}
\usepackage{MnSymbol}

\graphicspath{{img/}}

\numberwithin{equation}{section}

\newtheorem{theorem}{Theorem}[section]
\newtheorem{prop}[theorem]{Proposition}
\newtheorem{cor}[theorem]{Corollary}
\newtheorem{lemma}[theorem]{Lemma}
\newtheorem{definition}[theorem]{Definition}

\def\rq{\noindent\textbf{Remark : }}
\def\rqs{\noindent\textbf{Remarks : }}

\newcommand{\Mbar}{\overline{M}}

\newcommand{\bR}{\mathbb{R}}
\newcommand{\bB}{\mathbb{B}}
\newcommand{\bH}{\mathbb{H}}
\newcommand{\cC}{\mathcal{C}}
\newcommand{\cM}{\mathcal{M}}

\newcommand{\cG}{\mathcal{G}}

\newcommand{\gbar}{\overline{g}}
\def\into{\hookrightarrow}
\newcommand{\Wtil}{\widetilde{W}}
\newcommand{\Ttil}{\widetilde{T}}

\DeclareMathOperator{\tr}{tr}
\DeclareMathOperator{\divg}{div}

\DeclareMathOperator{\im}{Im}
\newcommand{\rlie}{\mathring{\mathcal{L}}}

\newcommand{\diff}  [2]{\frac{d #1}{d #2}}

\newcommand{\tg}{\widetilde{g}}

\newcommand{\scal}{\mathrm{Scal}}
\newcommand{\hscal}{\widehat{\scal}}

\newcommand{\ric}{\mathcal{R}\mathrm{ic}}

\newcommand{\ricdd}[2]{\ric_{#1 #2}}

\newcommand{\ricuu}[2]{\ric^{#1 #2}}

\newcommand{\riem}{\mathcal{R}}

\newcommand{\grad}[1]{\nabla_{#1}}

\begin{document}

\title{Linearization stability of the Einstein constraint equations on an asymptotically hyperbolic manifold}
\author{Romain Gicquaud}
\date{\today}
\maketitle

\begin{abstract}
We study the linearization stability of the Einstein constraint equations on an asymptotically hyperbolic manifold. In particular we prove that these equations are linearization stable in the neighborhood of vacuum solutions for a non-positive cosmological constant and of Friedman--Lema\^itre--Robertson--Walker spaces in a certain range of decays. We also prove that this result is no longer true for faster decays. The construction of the counterexamples is based on a new construction of TT-tensors on the Euclidean space and on positive energy theorems.
\end{abstract}

\tableofcontents

\pagebreak
\section{Introduction}
General Relativity describes space-time as a manifold $\cM^{n+1}$ endowed with a Lorentzian metric $\tg$, that is to say a symmetric 2-tensor with signature $(-++\cdots +)$, satisfying the Einstein's equations: 
\begin{equation} \label{eqEinstein}
\widetilde{G}_{\mu\nu} + \Lambda_c \tg_{\mu\nu} = \frac{8 \pi \cG}{c^4} \Ttil_{\mu\nu},
\end{equation} where $\widetilde{G}_{\mu\nu} = \ricdd{\mu\nu} - \frac{\hscal}{2} \tg_{\mu\nu}$ is the Einstein tensor of $\tg$, $\Lambda_c$ is the cosmological constant, $\cG$ is Newton's constant, $c$ is the speed of light and $\Ttil_{\mu\nu}$ is the stress-energy tensor of the matter. We refer the reader to any good reference textbook for more details (e.g. \cite{Wald} or \cite{Hawking}). These equations form an intricate system of non-linear partial differential equations and only few exact solutions are known. One is then lead to study perturbative solutions of the Einstein equations, that is to say solutions $(\cM, \tg')$ of the Einstein equations close to exact solution $(\cM, \tg)$. This issue is generally addressed by studying the Einstein equations linearized at $(\cM, \tg, \Ttil)$:
\begin{equation} \label{eqEinsteinLin}
D\widetilde{G}_{\tg} (\delta\tg) = \frac{8 \pi \cG}{c^4} \delta \Ttil,
\end{equation}
where $DG$ is the linearized Einstein operator, $\delta \tg = \tg' - \tg$ and $\delta \Ttil = \Ttil' - \Ttil$. As an example, this method is useful for the study of gravitational waves  (see e.g. \cite[section 4.4]{Wald}). It is then an important issue to know if all these approximate solutions of the Einstein equations can be converted into true solutions, i.e. given $\delta\tg$ and $\delta \Ttil$ satisfying the linearized Einstein equations \eqref{eqEinsteinLin}, does there exist a one-parameter pair $(\tg(\lambda), \Ttil(\lambda))$ solution of the Einstein equation \eqref{eqEinstein} such that $\tg(\lambda) = \tg + \lambda \delta \tg + o(\lambda)$ and $\Ttil(\lambda) = \Ttil + \lambda \delta \Ttil + o(\lambda)$ for a certain norm?\\

If the space-time $(\cM, \tg)$ is globally hyperbolic, this question is usually tackled in the following way. First remark that if $M \subset \cM$ is a Cauchy surface then $\delta \tg$ and $\delta \Ttil$ induce solutions of the linearized constraint equations on $M$. Then prove that the constraint equations are linearization stable (in a sense similar to the previous one, see definition \ref{defLinStab} for the precise statement). This problem turns out to be easier because it is elliptic. Finally prove that the solutions of the Einstein equations depends smoothly on the Cauchy data (see e.g. \cite{FriedrichRendall}).\\

This method was used to study the linearization stability of the vacuum Einstein equations when the space-time admits a compact Cauchy surface (see e.g. \cite{FischerMarsden}, \cite{Moncrief1, Moncrief2} and \cite{ArmsMarsden}) or an asymptotically Euclidean one \cite{CBD, CBFM}. Linearization stability of the Einstein equations coupled to matter has been addressed, among other, in the case of scalar-tensor theories in \cite{Saraykar1, Saraykar2} and of Friedmann--Lema\^itre--Robertson--Walker (FLRW) models (see \cite[sections 5.3 and 5.4]{Hawking} or \cite[section 5]{Wald} for the description of these models) in \cite{BrunaGirbau1, BrunaGirbau2, BrunaGirbau3}.\\

We extend results on linearization stability of the constraint equations to asymptotically hyperbolic Cauchy surfaces for vacuum space-times (Proposition \ref{propLinStabVide}) and FLRW models (Proposition \ref{propLinStabFLRW}). We prove that the constraint equations are linearization-stable in a certain weight interval (see Theorem \ref{thmLinStab} for the precise statment). We give counterexamples in Proposition \ref{propInstabAdS}, showing that the constraint equations are not garanteed to be linearization-stable for faster decay. We also apply the same procedure to the asymptotically Euclidean case in Proposition \ref{propInstabAE}. The construction of these counterexamples relies on positive energy theorems for asymptotically anti-de Sitter \cite{MaertenPositiveEnergy} and asymptotically flat \cite{ChruscielMaerten} space-times and on a new construction of compactly supported TT-tensors (Proposition \ref{propTTcompact}).\\

Starting from a pair $(\cM, \tg)$ and a Cauchy surface $M \subset \cM$, we denote $g$ the metric induced on $M$ by $\tg$ and $K$ the second fundamental form of the embedding $M \subset \cM$: let $N$ be the future pointing unit normal vector to $M$ and define (locally) the geodesic flow generated by $N$. This induces a time coordinate $t$ which is zero on $M$ and whose gradient is $-N$. Our convention for the second fundamental form is then the following:

\begin{equation} \label{defSecondeFormeFondamentaleLorentz}
K_{ij} = \frac{1}{2} \partial_t g_{ij} = \nabla_i N_j,
\end{equation}

where $i, k, \ldots$ correspond to spatial coordinates. The constraint equations then read:
\begin{eqnarray}
\label{contrainteHamilton} \scal_g - 2 \Lambda_c - \left|K\right|^2_g + \left(\tr_g K\right)^2 = 2 \rho
	& \qquad\text{(Hamiltonian constraint)}\\
\label{contrainteMoment}   \mathrm{div}_g K - d\left(\tr_g K\right) = - J_i & \qquad\text{(Momentum constraint)}
\end{eqnarray}
where $\left(\mathrm{div}_g K\right)_j = \nabla^i K_{ij}$, $\rho = \frac{8 \pi \cG}{c^4} T_{NN}$ and $J_i = \frac{8 \pi \cG}{c^4} T_{Ni}$. We refer the reader to \cite{BartnikIsenberg} or \cite{Gicquaud} for more details.\\

We shall denote $\Phi$ the constraint operator :

\begin{equation}
\Phi(g, K)
 	= \left(\begin{array}{l}
		\scal_g - 2 \Lambda_c - \left|K \right|^2_g + \left(\tr_g K\right)^2\\
		\nabla^j K_{ij} - \nabla_i \left(\tr_g K\right)
	\end{array} \right)
	= \left(\begin{array}{l}
		2 \rho\\
		-J_i
	\end{array} \right).
\end{equation}

If the pair $(g, K)$ in the neighborhood of which we are willing to study the linearization stability of the constraint equations is not in the functional space we are considering (which will be the case for function spaces in which the variations of the metric tends to 0 at infinity), we have to consider affine spaces so it is convenient to introduce the following renormalized constraint operator:

\begin{equation}\label{eqOpContrainteRen}
\Phi_r : (\delta g, \delta K) \mapsto \Phi(g + \delta g, K + \delta K) - \Phi(g, K).
\end{equation}

This paper is organized as follows. In Section \ref{secVarAH}, we introduce the class of asymptotically hyperbolic manifolds together with the functional spaces naturally attached to them. In Section \ref{secLinStab}, we first give the precise definition of linearization stability (Subsection \ref{secDefLinStab}). Then we give conditions for the differentiability of the constraint operator in Subsection \ref{secDiffContraintes}. We state and prove our main theorem (Theorem \ref{thmLinStab}) in Subsection \ref{secLinStabThm}. In Subsection \ref{secLinStabExemples}, we give two examples of spaces that are linearization stable (Propositions \ref{propLinStabVide} and \ref{propLinStabFLRW}). Finally, we give simple counterexamples of linearization instability in Subsection \ref{secLinStabCtrEx}.\\

\noindent\textit{Acknowledgments.} I am grateful to Gilles Carron and Erwann Delay for useful discussions and support. I also thank Piotr Chru\'sciel and Daniel Maerten for useful references.

\section{Asymptotically hyperbolic manifolds}\label{secVarAH}

\subsection{Definition}
Let $\Mbar$ be a smooth compact manifold with boundary $\partial M$. We denote $M$ the interior of $\Mbar$. A \textbf{defining function} for $\partial M$ is a smooth function $\rho : \Mbar \to [0;~\infty)$ such that $\rho^{-1}(0) = \partial M$ and $d\rho \neq 0$ along $\partial M$. A Riemannian metric $g$ on $M$ is called $\mathcal{C}^{l, \beta}$-\textbf{conformally compact} if $\rho^2 g$ extends to a $\mathcal{C}^{l, \beta}$ metric $\gbar$ on $\Mbar$. A simple calculation proves that if $g$ is $\cC^{l, \beta}$-conformally compact with $l + \beta \geq 2$ then the sectional curvature of $g$ tends to $-|d\rho|^2_{\gbar}$ in a neighborhood of $\partial M$. As a consequence, $g$ is said \textbf{asymptotically hyperbolic} if $g$ is conformally compact and such that $|d\rho|^2_{\gbar} = 1$ along $\partial M$.

\subsection{Function spaces}
Fix a geometric tensor bundle (i.e. associated to the principal $SO(M)$-bundle) E on $M$. Let $k \geq 0$ be an integer and $0 < \alpha < 1$. We first define the (non weighted) Sobolev space $W^{k, p}_0(M, E)$ as the set of sections $u \in L^p$ of $E$ such that $\forall~j\in \{0, \cdots, k\},~\nabla^{(j)} u \in L^p$ endowed with the norm:

$$\|u\|_{W^{k,p}_0(M, E)} = \left( \sum_{j=0}^k \int_M \left\|\nabla^{(j)} u\right\|^p d\mu_g \right)^{\frac{1}{p}}.$$

We also define the weighted Sobolev space $W^{k, p}_\delta (M, E) = \rho^\delta W^{k, p}_0 (M, E)$ together with the norm: $\|u\|_{W^{k,p}_\delta(M, E)} = \|\rho^{-\delta} u\|_{W^{k,p}_0(M, E)}$. We shall denote $L^p_\delta = W^{0, p}_\delta$.\\

Then we define weighted H\"older spaces. Choose a finite number of charts $\phi = (\rho, \theta^1, \cdots, \theta^{n-1})$ in the neighborhood of $\partial M$ such that the reunion of their domain of definition covers $\partial M$. Add another finite set of charts defined on relatively compact domains to cover the whole of $M$. Denote $\mathbb{H}^n$ the upper half-space model of the hyperbolic space : $\{(x_1, \ldots, x_n) \in \mathbb{R}^n \vert x_1 > 0 \}$ endowed with the metric $g_{hyp} = \frac{1}{x_1^2} g_{eucl}$. Define $B_r$ the ball of radius $r$ centered at $(1,0, \cdots, 0)$ in $\mathbb{H}$ with the hyperbolic metric. If $M \ni p_0 = \phi^{-1}(\rho_0, \theta^1_0\cdots, \theta^{n-1}_0)$ is the reciprocal image for one of the given charts, define the (radius $r$-) M\"obius chart at $p_0$ by :

$$\phi^r_{p_0}\left(p\right) = \left(\frac{\rho(p)}{\rho_0}, \frac{\theta^1(p)-\theta^1_0}{\rho_0}, \cdots, \frac{\theta^{n-1}(p)-\theta^{n-1}_0}{\rho_0} \right)$$

$\phi^r_{p_0} : \left(\phi^r_{p_0}\right)^{-1}\left(B_r\right) \to B_r$. The H\"older norm is defined by :

$$\|u\|_{\mathcal{C}^{k, \alpha}_\delta(M, E)} = \sup_{p_0 \in M} \rho^{-\delta}(p_0) \left\| \left(\left(\phi^1_{p_0}\right)^{-1}\right)^* u\right\|_{\mathcal{C}^{k, \alpha}(B_1)}.$$

The space $\mathcal{C}^{k, \alpha}_\delta(M, E)$ is the set of sections $u \in \mathcal{C}^{k, \alpha}_{\mathrm{loc}}$ such that :

$$\|u\|_{\mathcal{C}^{k, \alpha}_\delta(M, E)} < \infty.$$

For more details about these function spaces, we refer the reader to \cite{LeeFredholm}. An important remark is that, unlike the asymptotically Euclidean case (see e.g. \cite{Bartnik}), there is no gain of decay in the Sobolev injections. We give a simple example of this fact : There is no continuous injection $W^{1, p}_\delta(M, \bR) \into L^q_{\delta'}(M, \bR)$ for any $\delta' > \delta$. Indeed, let $M, g$ be any AH manifold. Choose a smooth compactly supported function $f : B_1 \to \mathbb{R}$. Select a sequence of M\"obius balls $B_i$ tending to infinity, set $f_i = f \circ \Phi_i$ where $\Phi_i$ is the chart associated to $B_i$ and extend $f$ by zero outside $B_i$. It can be easily seen that the sequence of functions $\frac{f_i}{\rho_i^\delta}$ is bounded in $W^{1, p}_\delta(M, \bR)$ but diverges in $L^q_{\delta'}(M, \bR)$ if $\delta' > \delta$.\\

We remark the following proposition whose proof is similar to \cite[Theorem 5.23]{Adams}:

\begin{prop} Let $(M^n, g)$ be a $\cC^{l, \beta}$-asymptotically hyperbolic manifold with $l+\beta \geq 2$, $1 \geq k \geq l$ an integer, and $1 \leq p \leq \infty$ a real number such that $k p > n$. The $W^{k, p}_0(M, \bR)$ space is a Banach algebra. The spaces $W^{k, p}_\delta(M, V)$, for $V$ any geometric bundle over $M$ are $W^{k, p}_0(M, \bR)$-modules and the map $W^{k, p}_0(M, \bR) \times W^{k, p}_\delta(M, V) \to W^{k, p}_\delta(M, V)$ is continuous.
\end{prop}

This propositions generalizes to the spaces $W^{k, p}_\delta(M, \bR)$ for $\delta \geq 0$: the map $W^{k, p}_\delta(M, \bR) \times W^{k, p}_{\delta'}(M, V) \to W^{k, p}_{\delta+\delta'}(M, V)$ is bilinear continuous. In particular, the spaces $W^{k, p}_\delta(M, \bR)$ are Banach algebras (without identity) for $\delta > 0$. The previous counter-example proves that this property is no longer true when $\delta < 0$. This makes Sobolev spaces not suited to the study of non-linear problems on asymptotically hyperbolic manifolds because it is not possible to obtain the full expected weight range (as an example, one would expect naively that the condition $\delta < \delta_0$ for H\"older spaces transforms into $\delta + \frac{n-1}{p} < \delta_0$ for Sobolev spaces as it is the case for linear equations). We give another example: if $g$ is an asymptotically hyperbolic metric, the set of positive definite metrics is an open subset of the affine space $g + W^{k, p}_\delta$ only if $\delta \geq 0$. However, it is possible to get these properties back by restricting ourselves to subspaces of $W^{k, p}_\delta$. We define the $\Wtil^{m, p; k, \alpha}_\delta$ spaces:

\begin{equation}
\Wtil^{m, p; k, \alpha}_\delta (M, V) = W^{m, p}_\delta(M, V) \bigcap \cC^{k, \alpha}_0(M, V).
\end{equation}

where $0 \leq m, k \leq l+\beta$ are integers, $1 \leq p < \infty$, $0 \leq \alpha < 1$ with $k + \alpha \leq l + \beta$. Remark that this intersection makes sense because both these spaces are subspaces of $L^p_{\delta'}$ for some $\delta' < 0$ large enough. We endow this space with the norm:

\begin{equation}
\left\| u \right\|_{\Wtil^{m, p; k, \alpha}_\delta (M, V)} = \left\|u\right\|_{W^{m, p}_\delta(M, V)} + \left\|u\right\|_{\cC^{k, \alpha}_0(M, V)}.
\end{equation}

We first show that these spaces behave well with respect to tensor product:

\begin{prop}\label{propWtilBanach} Let $(M, g)$ be a $\cC^{l, \beta}$-asymptotically hyperbolic manifold. The space $\Wtil^{m, p; k, \alpha}_\delta (M, V)$ is a Banach space. If $k + \alpha \geq \lfloor \frac{m}{2} \rfloor$, with $m \leq l+\beta$ and $k+\alpha \leq l+\beta$, for all $\delta \in \bR$, the product map $\Wtil^{m, p; k, \alpha}_\delta (M, V_1) \times \Wtil^{m, p; k, \alpha}_\delta (M, V_2) \to \Wtil^{m, p; k, \alpha}_\delta (M, V_1 \otimes V_2)$ is a continuous bilinear map.
\end{prop}

\begin{proof} We first prove that these spaces are Banach spaces. This is a consequence of the general following fact: if $X$, $Y$ and $Z$ are three Banach spaces, such that $X, Y$ are vector subspaces of $Z$, and such that the norms $\|.\|_X$ and $\|.\|_Y$ are stronger than the norms induced on $X$ and $Y$ by $\|.\|_Z$, the space $X \bigcap Y$ endowed with the norm $\|.\|_X + \|.\|_Y$ is a Banach space. Indeed if $\left(x_i\right)_{i \geq 0}$ is a Cauchy sequence in $X \bigcap Y$, then it is also a Cauchy sequence in $X$, $Y$ and $Z$. It admits a limit in all those three spaces. Since the norm on $Z$ is weaker than the norms on $X$ and $Y$, $\lim_X x_i = \lim_Z x_i = \lim_Y x_i \in X \bigcap Y$.\\

To prove the second property, we cover $M$ by a countable number of M\"obius charts $\left(B_i, \Phi_i\right)$, with $B_i$ centered at $x_i \in M$, such that there exists an $N < \infty$ such that each $B_i$ intersects at most $N$ $B_j$, $j \neq i$ (the existence of such a covering is shown in \cite{LeeFredholm}). We define the following norm:
$$
\left\|u\right\|_{W^{m, p}_\delta}' = \sum_i \rho^{-\delta}(x_i) \left\| \left(\Phi_i\right)_* u \right\|_{W^{m, p}(B)}
$$
where the $W^{k, p}$ norm appearing on the right hand side of this equation is the usual norm associated to the Euclidean metric on $B$. This norm is equivalent to $\left\|.\right\|_{W^{m, p}_\delta}$ we defined previously. The subspace $\cC^{l, \beta}(B)$ is dense in $W^{m, p}(B) \bigcap \cC^{k, \alpha}(B)$. Let $a = (a_1, \ldots, a_n)$ be an n-uplet of positive integers, define $|a| = \sum_i a_i$. If $a$ and $b$ are two n-uplets of positive integers, we say that $b \leq a$ if $b_i \leq a_i$ for all $i = 1, \ldots, n$. Let $u, v \in \cC^{l, \beta}(B)$ and $a$ be an n-uplet of positive integers such that $|a| \leq m$. One has : $$ \partial^a (u \otimes v) = \sum_{b \leq a} \partial^a u \otimes \partial^{a-b} v, $$ so
\begin{eqnarray*}
\left\|\partial^a (uv)\right\|_{L^p(B)}
	& \leq & \sum_{b \leq a} \left\|\partial^b u \otimes \partial^{a-b} v\right\|_{L^p(B)}\\
	& \leq & \sum_{b \leq a, |b| \leq \left\lfloor\frac{|a|}{2}\right\rfloor} \left\|\partial^b u\right\|_{L^\infty(B)} \left\|\partial^{a-b} v\right\|_{L^p(B)}
				 + \sum_{b \leq a, |b| > \left\lfloor\frac{|a|}{2}\right\rfloor} \left\|\partial^b u\right\|_{L^p(B)} \left\|\partial^{a-b} v\right\|_{L^\infty(B)}\\
	& \leq & \left\| u \right\|_{C^{k, \alpha}(B)} \left\| v \right\|_{W^{m,p}(B)} + \left\| v \right\|_{C^{k, \alpha}(B)} \left\| u \right\|_{W^{m,p}(B)}.
\end{eqnarray*}
This proves that
$$ \left\|u \otimes v \right\|_{W^{k, p}(B)} \leq C \left\| u \right\|_{C^{k, \alpha}(B)} \left\| v \right\|_{W^{m,p}(B)} + \left\| v \right\|_{C^{k, \alpha}(B)} \left\| u \right\|_{W^{m,p}(B)}. $$
for some constant $C > 0$. By density of the functions $\cC^{l, \beta}(B)$, this inequality holds for any $u, v \in W^{m, p}(B) \bigcap \cC^{k, \alpha}(B)$. As a consequence:
\begin{eqnarray*}
\left\|u \otimes v\right\|_{W^{m, p}_\delta(M, V_1 \otimes V_2)}'
	&   =  & \sum_i \rho^{-\delta}(x_i) \left\| \left(\Phi_i\right)_* \left(u \otimes v\right) \right\|_{W^{m, p}(B)}\\
	& \leq & C \sum_i \rho^{-\delta}(x_i) \left(\left\| \left(\Phi_i\right)_* u \right\|_{W^{m, p}(B)} \left\| \left(\Phi_i\right)_* v \right\|_{\cC^{k, \alpha}(B)}\right.\\
	&		 & + \left.\left\| \left(\Phi_i\right)_* u \right\|_{\cC^{k, \alpha}(B)} \left\| \left(\Phi_i\right)_* v \right\|_{W^{m, p}(B)} \right)\\
	& \leq & C \left(\left\|u\right\|_{\cC^{k, \alpha}_0(M, V_1)}\left\|v\right\|_{W^{m, p}_\delta(M, V_2)}' + \left\|u\right\|_{W^{m, p}_\delta(M, V_1)} \left\|v\right\|_{\cC^{k, \alpha}_0(M, V_2)}'\right)\\
\left\|u \otimes v\right\|_{W^{m, p}_\delta(M, V_1 \otimes V_2)}
	& \leq & C' \left\|u\right\|_{\Wtil^{m, p; k, \alpha}_\delta(M, V_1)} \left\|v\right\|_{\Wtil^{m, p; k, \alpha}_\delta(M, V_2)},
\end{eqnarray*}
for some constant $C' > 0$ independant of $u$ and $v$. We also have the inequality:
$$
\left\|u \otimes v\right\|_{\cC^{k, \alpha}(M, V_1 \otimes V_2)}
	\leq C'' \left\|u\right\|_{\cC^{k, \alpha}_0(M, V_1)} \left\|v\right\|_{\cC^{k, \alpha}_0(M, V_2)}
	\leq C'' \left\|u\right\|_{\Wtil^{m, p; k, \alpha}_\delta(M, V_1)} \left\|v\right\|_{\Wtil^{m, p; k, \alpha}_\delta(M, V_2)},
$$
where $C'' > 0$ is independant of $u$ and $v$. The Proposition is then obtained by adding the two inequalities.
\end{proof}

\begin{cor}\label{propWtilInverse} Under the assumptions of Proposition \ref{propWtilBanach}, the map $g' \mapsto (g')^{-1}$ is well defined and analytic in a neighborhood of $g$ between the following affine spaces: $$g + \Wtil^{m, p; k, \alpha}_\delta \to g^{-1} + \Wtil^{m, p; k, \alpha}_\delta.$$
\end{cor}

\begin{proof}
By raising one index, this amounts to showing that the map $$Id + \Wtil^{m, p; k, \alpha}_\delta(End(TM)) \to Id + \Wtil^{m, p; k, \alpha}_\delta(End(TM))$$ which associates to a section of the bundle of endomorphism of the tangent bundle its inverse is well defined and analytic in a neighborhood of $Id$. To prove this, we remark that: the space $\Wtil^{m, p; k, \alpha}_\delta(End(TM))$ is a Banach algebra and there is a continuous injection $\Wtil^{m, p; k, \alpha}_\delta \into L^\infty$. Indeed if $W$ is a Banach algebra which is continuously embedded in $L^\infty$, the application $Id + u \mapsto (Id + u)^{-1}$, with $u \in W$, is given by:
$$(Id + u)^{-1} = Id - u + u^2 - u^3 + \ldots$$ This series converges in $L^\infty$ if $\left\|u\right\|_{L^\infty} < 1$ and also converges in $Id + W$ if $\left\|u\right\|_{W}$ is small enough. The injection $W \into L^\infty$ begin continuous, the limits of these series coincides. This proves the corollary.
\end{proof}

\section{Linearization stability}\label{secLinStab}

\subsection{Preliminaries on Banach spaces}\label{secDefLinStab}
In this section, we give the general definition of linearization stability for a non-linear function $F$ between two Banach spaces (definition \ref{defLinStab}) and give a practical criterion to prove linearization stability (Proposition \ref{propLinStab2}). The definition we give is rather weak but turns out to be sufficient in practice. We refer the reader to \cite{Moncrief1} or \cite{BrunaGirbau1} for stronger definitions.

\begin{definition}\label{defLinStab} Let $X$ and $Y$ be two Banach spaces, $U$ an open subset of $X$ and $F : U \to Y$ a continuous map. Let $x_0 \in U$ and denote $y_0 = F(x_0)$. $F$ is said \textbf{linearization stable} at $x_0$ if $F$ is differentiable at $x_0$ and if for any $\delta x \in X$ and any $\delta y \in Y$ such that $\delta y = DF_{x_0} (\delta x)$, there exist $\epsilon > 0$ and a curve $x = x(\lambda) \in X,~\lambda \in (-\epsilon, \epsilon)$ such that:
$$
\left\lbrace
\begin{aligned}
F(x) & = y_0 + \lambda \delta y\\
x    & = x_0 + \lambda \delta x + o(\lambda).
\end{aligned}
\right.
$$
\end{definition}

\rqs
\begin{enumerate}
\item In the definition, we have chosen $y = y_0 + \lambda \delta y$ while the definition given in the introduction corresponds to $y = y_0 + \lambda \delta y + o(\lambda)$. This is due to the fact that constraint equations do not restrict the choice of $\Ttil$ along $M$ while Einstein's equations imply the conservation of the stress-momentum tensor: $\widetilde{\nabla}^\mu \Ttil_{\mu\nu} = 0$ (which is a consequence of the motion equations for all fields but the gravitational one).
\item The curve $x(\lambda)$ is not (locally) unique in general (unless $DF_{x_0}$ is an isomorphism).
\end{enumerate}

\begin{prop}\label{propLinStab} Let $X$ and $Y$ be two Banach spaces, $U$ an open subset of $X$ and $F : U \to Y$ a continuous map. Let $x_0 \in U$ be such that $F$ is differentiable at $x_0$. If $D F_{x_0}$ is onto and if $\ker \left( D F_{x_0}\right)$ admits a closed complementary subspace then $F$ is linearization stable at $x_0$.
\end{prop}

\begin{proof}
Let $Z$ be a closed complementary subspace of $\ker \left( D F_{x_0}\right)$ in $X$. Without loss of generality, we can assume that $x_0, y_0 = 0$. The map $D F_{0}$ restricted to $Z$ is a bicontinuous isomorphism. Let $p$ be the projection onto $\ker \left( D F_0\right)$ with respect to $Z$. The map $\tilde{F} : U \to Y \times \ker \left( D F_0\right)$, $x \mapsto (F(x), p(x))$ is continuous and differentiable at $0$ with $D \tilde{F}_0 : \delta x \mapsto (D F_0(\delta x), p(\delta x))$. $D \tilde{F}_0$ is an isomorphism. The Implicit Function Theorem shows that there exists neighborhoods $V_0$ of $0_X$, $V_1$ of $0_Y$ and a function $\tilde{F}^{-1} : V_1 \to V_0$ reciprocal to $F$. Let $x(\lambda) = \tilde{F}^{-1}(D \tilde{F}_0(\lambda \delta x))$. $x(\lambda)$ is the unique solution in $V_0$ of $F(x) = \lambda \delta y,~p(x) = \lambda \delta x$. By the definition of the differential of $F$, one has $$x(\lambda) - \lambda \delta x = \tilde{F}^{-1}(D \tilde{F}_0(\lambda \delta x)) - D\tilde{F}^{-1}_0(D \tilde{F}_0(\lambda \delta x)) = o(\lambda).$$
\end{proof}

We give now a simple criterion to prove linearization stability:

\begin{prop}\label{propLinStab2} Under the assumptions of Proposition \ref{propLinStab}, if there exists a Banach space $E$ and a linear map $\varphi : E \to X$ such that $D F_{x_0} \circ \varphi : E \to Y$ is an isomorphism, then $F$ is linearization stable at $x_0$.
\end{prop}

\begin{proof} From Proposition \ref{propLinStab}, it is enough to prove that $D F_{x_0}$ is onto and that $\ker D F_{x_0}$ admits a closed complementary subspace. It is obvious that $D F_{x_0}$ is onto and that $\im \varphi \oplus \ker D F_{x_0} = X$. There is only to prove that $\im \varphi$ is closed in $X$. There exists a constant $C > 0$ such that $\forall x \in E~\|x \|\leq C \|\varphi(x)\|$. Indeed, there exists $C' > 0$ such that $\| x \| \leq C' \| D F_{x_0} (\varphi(x))\|$ so $\| x \| \leq C' \left\| D F_{x_0} \right\|~\|\varphi(x)\|$. This proves that $\varphi : E \to \im \varphi$ is a bicontinuous morphism. In particular $\im \varphi$ is complete for the norm $\|.\|_X$ and so it is closed.
\end{proof}

Remark in particular that the linearization stability issue is closely related to the study of the manifold structure of the set of solutions of the constraint equations. This problem is studied in \cite{ChruscielDelayManifold}.

\subsection{Differentiability of the constraint equations}\label{secDiffContraintes}
\begin{prop}[Differentiability of the constraint operator]
Let $(M, g)$ be a $\cC^{l, \beta}$-asymptotically hyperbolic manifold, with $l+\beta > 2$, and $K$ a symmetric 2-tensor, then the renormalized constraint operator $$\Phi_r : (\delta g, \delta K) \mapsto \Phi(g + \delta g, K + \delta K) - \Phi(g, K) = (2 \rho(g + \delta g) - 2 \rho(g), -J_i(g + \delta g) + J_i(g))$$ is well defined in a neighborhood of $(0, 0)$ and is differentiable at $(0, 0)$ seen as an operator between the following spaces:
$$
\left\lbrace
\begin{array}{rll}
\cC^{m+2, \alpha}_\delta \times \cC^{m+1, \alpha}_\delta & \to \cC^{m, \alpha}_\delta   \times \cC^{m, \alpha}_\delta
	&\text{if $\delta \geq 0$ and $m+2+\alpha \leq l + \beta$}\\
W^{m+2, p}_{\delta} \times W^{m+1, p}_{\delta} & \to W^{m, p}_{\delta} \times W^{m, p}_{\delta}
	&\text{if $\delta \geq 0$, $n < m p$ and $m+2\leq l + \beta$}\\
\Wtil^{m+2,p; k+2, \alpha}_\delta \times \Wtil^{m+1,p; k+1, \alpha}_\delta & \to \Wtil^{m,p; k, \alpha}_\delta \times \Wtil^{m,p; k, \alpha}_\delta
	&\text{if $m \leq l + \beta$ and $\lfloor \frac{m}{2} \rfloor \leq k + \alpha \leq l + \beta$,}\\
&	&\text{for all $\delta \in \bR$.}
\end{array}
\right.
$$

The linearized constraint operator at $(g, K)$ is then given by:
\begin{equation}\label{eqContrainteLin}
\varphi(h, k) =
\left(
\begin{array}{c}
-\Delta \left(\tr_g h\right) + \grad{p}\grad{q} h^{pq} - \ricuu{p}{q} h_{pq} - 2 K^{pq} k_{pq}\\
\quad + 2 g^{kl} g^{pq} g^{uv} K_{kp} K_{qu} h_{vq} + 2 \left(\tr_g K\right) g^{pq} \left(k_{pq} - h_{pk} g^{kl} K_{lq}\right)\\
 \\
\nabla^j k_{ij} - \nabla_i \left(\tr_g k\right) - h^{pq} \nabla_p K_{iq} + \frac{1}{2} K^{pq} \nabla_i h_{pq} + h^{pq} \nabla_i K_{pq}\\
\quad - \frac{1}{2} g^{pq} \left[2 \nabla^k h_{kp} - \nabla_p \left(\tr_g h\right) \right] K_{iq}
\end{array}
\right)
\end{equation}
where we set $h = \delta g$ and $k = \delta K$ the variations of the metric and of the second fundamental form.
\end{prop}

\begin{proof} The proof is the same for all three cases. For H\"older we also refer the reader to \cite{ChruscielDelayMapping}. We show the differentiability of the contraint map between the $\Wtil$ spaces. We will just prove the differentiability of the map $\delta g \mapsto \scal_{g + \delta g} - \scal_g$, the proof for the other terms appearing in the definition of $\Phi_r$ being similar. Denote $g' = g + \delta g$. The difference between the Levi-Civita connexions associated to $g$ and to $g'$ is a tensor: $$\nabla' = \nabla + \Gamma,$$ where the expression of $\Gamma$ is identical to the usual expression of the Christoffel symbols (for the metric $g'$) except that the partial derivatives are replaced by covariant derivatives (with respect to $g$). The (1, 3) Riemann tensor associated to $g'$ can then be written formally: $$\riem_{g'} = \riem_g + \nabla \Gamma - \nabla \Gamma + \Gamma * \Gamma - \Gamma * \Gamma.$$ In particular, the scalar curvature of $g'$ is $$\scal_{g'} = (g')^{-1}*\ric_g + (g')^{-1}*\left(\nabla \Gamma - \nabla \Gamma + \Gamma * \Gamma - \Gamma * \Gamma\right).$$ The differentiability of $g' \mapsto \scal_{g'}$ is then a simple consequence of the stability with respect to tensor product of the spaces $\Wtil^{m, p; k, \alpha}_\delta$ (Proposition \ref{propWtilBanach}) which implies that the product of two differentiable maps is differentiable and of its corollary \ref{propWtilInverse} (diff\'erentiability of the inverse of the metric).
\end{proof}

\subsection{Linearization stability of the constraint equations}\label{secLinStabThm}
Let $(M, g)$ be a $\cC^{l, \beta}$-asymptoticaly hyperbolic manifold. Let $m \geq 0$, $\alpha \in [0; 1)$ be such that $m + 2 + \alpha \leq l + \beta$. Define $h = u g$ and $m = \rlie_\xi g + \alpha u K + \beta u \left(\tr_g K\right) g$. If $K \in \cC^{m+1, \alpha}_\delta$, the map $$f : (u, \xi) \to (h, k)$$ is a linear map between the following spaces:

$$
\left\lbrace
\begin{array}{rl}
\cC^{m+2, \alpha}_\delta \times \cC^{m+2, \alpha}_\delta & \to \cC^{m+2, \alpha}_\delta   \times \cC^{m+1, \alpha}_\delta\\
W^{m+2, p}_{\delta} \times W^{m+2, p}_{\delta} & \to W^{m+2, p}_{\delta} \times W^{m+1, p}_{\delta}\\
\Wtil^{m+2, p; k+2, \alpha}_{\delta} \times \Wtil^{m+2, p; k+2, \alpha}_{\delta} & \to \Wtil^{m+2, p; k+2, \alpha}_{\delta} \times \Wtil^{m+1, p; k+1, \alpha}_{\delta}
\end{array}
\right.
$$

Composition with the linearized constraint operator $\varphi$ \eqref{eqContrainteLin} leads to:

\begin{eqnarray*}
\varphi \circ f(u, \xi)
	& = & \left(
\begin{array}{c}
-\Delta \left(\tr_g h\right) + \grad{p}\grad{q} h^{pq} - \ricuu{p}{q} h_{pq} - 2 K^{pq} k_{pq}\\
\quad + 2 g^{kl} g^{pq} g^{uv} K_{kp} K_{qu} h_{vq} + 2 \left(\tr_g K\right) g^{pq} \left(k_{pq} - h_{pk} g^{kl} K_{lq}\right)\\
 \\
\nabla^j k_{ij} - \nabla_i \left(\tr_g k\right) - h^{pq} \nabla_p K_{iq} + \frac{1}{2} K^{pq} \nabla_i h_{pq} + h^{pq} \nabla_i K_{pq}\\
\quad - \frac{1}{2} g^{pq} \left[2 \nabla^k h_{kp} - \nabla_p \left(\tr_g h\right) \right] K_{iq}
\end{array}
\right)\\
	& = & \left(
\begin{array}{c}
- (n-1) \Delta u - \scal~u + 2 \left| K\right|^2_g u - 2 \left(\tr_g K\right)^2 u - 2 \left(K^{pq} - \left(\tr_g K\right) g^{pq} \right) k_{pq}\\
 \\
\nabla^j k_{ij} - \nabla_i \left(\tr_g k\right) - u \nabla^j K_{ij} + \frac{1}{2} \left(\tr_g K\right) \nabla_i u + u \nabla_i \left(\tr_g K\right)
+ \frac{n-2}{2} K_{ij} \nabla^j u
\end{array}
\right)
\end{eqnarray*}

$$
\varphi \circ f(u, \xi) = \left(
\begin{array}{c}
- (n-1) \Delta u - \scal~u + 2 \left| K\right|^2_g u - 2 \left(\tr_g K\right)^2 u\\
- 2 \left[\left\langle K, \rlie_\xi g \right\rangle +\alpha u \left(\left|K\right|_g^2 - \left(\tr_g K\right)^2\right) - (n-1) \beta u \left(\tr_g K\right)^2\right]\\
 \\
\Delta_{TT} \xi_i + (\alpha-1) u \left(\nabla^j K_{ij} - \nabla_i \left(\tr_g K\right)\right) + \alpha\left[K_{ij} \nabla^j u-\left(\tr_g K\right) \nabla_i u \right]\\
- (n-1) \beta \left[u \nabla_i \left(\tr_g K\right) + \left(\tr_g K\right) \nabla_i u\right] + \frac{1}{2} \left(\tr_g K\right) \nabla_i u + \frac{n-2}{2} K_{ij} \nabla^j u
\end{array}
\right)
$$

Choosing $\alpha = - \frac{n-2}{2}$ and $\beta = \frac{1}{2}$, we obtain:
\begin{equation}\label{eqDefInverseContrainte}
f(u, \xi) = \left(
\begin{array}{c}
u g\\
\rlie_\xi g - \frac{n-2}{2} u K + \frac{1}{2} u \left(\tr_g K\right) g
\end{array}
\right)
\end{equation}
and
$$
\varphi \circ f(u, \xi) = \left(
\begin{array}{c}
- (n-1) \Delta u - \scal~u + n \left(\left| K\right|^2_g - \left(\tr_g K\right)^2\right) u
+ (n-1) u \left(\tr_g K\right)^2 - 2 \left\langle K, \rlie_\xi g \right\rangle\\
 \\
\Delta_{TT} \xi_i - \frac{n}{2} u \left(\nabla^j K_{ij} - \nabla_i \left(\tr_g K\right)\right) - \frac{n-1}{2} u \nabla_i \left(\tr_g K\right)
\end{array}
\right)
$$
\begin{equation}
\varphi \circ f(u, \xi) = \left(
\begin{array}{c}
- (n-1) \Delta u - \scal~u + n \left| L\right|^2_g u - 2 \left\langle L, \rlie_\xi g \right\rangle\\
 \\
\Delta_{TT} \xi_i + \frac{n}{2} u J_i - \frac{n-1}{2} u \nabla_i \left(\tr_g K\right)
\end{array}
\right),
\end{equation}
where $L = \mathring{K}$ is the traceless part of the second fundamental form $K$, $J_i = \nabla_i \left(\tr_g K\right) - \nabla^j K_{ij}$ and $\Delta_{TT} \xi_i = \divg \left(\rlie_\xi g\right)_i = \Delta \xi_i + \nabla^j\nabla_i\xi_j - \frac{2}{n} \nabla_i \left(\nabla^j\xi_j\right)$.\\

We give the following theorem which is a corollary of the proof of \cite[Theorem 1.3]{Gicquaud}:

\begin{lemma}[Isomorphism Theorem for $\Delta_{TT}$]\label{lmIsoTT} Let $(M, g)$ be a $\cC^{l, \beta}$-asymptotically hyperbolic manifold with $l+\beta \geq 2$. The linear map $\Delta_{TT}$ is an isomorphism between the following spaces:
$$
\left\lbrace
\begin{array}{rll}
\cC^{m+2, \alpha}_\delta & \to \cC^{m, \alpha}_\delta & \text{if $m+2+\alpha \leq l+\beta$ and $\delta \in (-1, n)$}\\
W^{m+2, p}_{\delta}			 & \to W^{m, p}_{\delta} & \text{if $m+2\leq l+\beta$ and $\left|\delta+\frac{n-1}{p}-\frac{n-1}{2}\right| < \frac{n+1}{2}$}.
\end{array}
\right.
$$
\end{lemma}

\begin{lemma}[Isomorphism Theorem for the linearized Hamiltonian constraint]\label{lmIsoHam} Let $(M, g)$ be a $\cC^{l, \beta}$-asymptotically hyperbolic manifold and $L$ be a symmetric traceless 2-tensor such that $\left|L\right|^2 \in \cC^{0, \alpha}_{\delta'}$ for a certain $\delta' > 0$. If the operator $$P : u \mapsto -(n-1) \Delta u - \scal~u + n \left|L\right|^2 u$$ has a trivial $L^2$-kernel, then it is an isomorphism between the following spaces:
$$
\left\lbrace
\begin{array}{rll}
\cC^{m+2, \alpha}_\delta & \to \cC^{m, \alpha}_\delta &\text{if $\delta \in (-1, n)$ and $L \in \cC^{m, \alpha}_0$}\\
W^{m+2, p}_{\delta}			 & \to W^{m, p}_{\delta} &\text{if $\left|\delta + \frac{n-1}{p} - \frac{n-1}{2} \right| < \frac{n+1}{2}$ and $L \in \cC^{m,0}_0$}.
\end{array}
\right.
$$
\end{lemma}

\rq This isomorphism Theorem is valid for Sobolev spaces assuming only that $\left|L\right|^2 \to 0$ at infinity.

\begin{proof} This theorem cannot be proved by using \cite[Theorem C]{LeeFredholm} because this Theorem only applies to geometric operators (see also \cite{AnderssonChrusciel}). Even if it is possible to modify the proof of this Theorem, we give another one which uses \cite[Theorem C]{LeeFredholm} without modification. We deal only with Sobolev spaces. The proof for H\"older spaces is similar (but slightly more complicated). Assume first that $m = 0$. Define $P_0 : u \mapsto -(n-1) \Delta u - \scal~u$. $P_0$ is a formally selfadjoint geometric operator. A straightforward calculation shows that the critical exponents of $P_0$ are $s = -1$ and $s = n$. Let $p \in (1, \infty)$ and $\delta$ be such that $\left|\delta + \frac{n-1}{p} - \frac{n-1}{2} \right| < \frac{n+1}{2}$. By \cite[Theorem C]{LeeFredholm}, the operator $$P_0 : W^{2, p}_\delta \to L^p_\delta$$ is Fredholm with zero index. Let $\psi : [0, \infty) \to \bR$ be a cut-off function such that $\psi(x) = 1$ when $0 \leq x \leq 1$ and $\psi(x) = 0$ when $x \geq 2$. Let $\epsilon > 0$, define the operator $P_\epsilon : W^{2, p}_\delta \to L^p_\delta$ by $P_\epsilon(u) = -(n-1) \Delta u - \scal~u + n \left|L\right|^2 \psi\left(\frac{\rho}{\epsilon}\right) u$. As $\left|L\right|^2 \to 0$ at infinity, the operators $P_\epsilon$ converge to $P_0$ as operators from $W^{2, p}_\delta$ to $L^p_\delta$. If $\epsilon$ is small enough, the operator $P_\epsilon$ is Fredholm with index $0$. If $\delta' < \delta$, there exists a constant $C > 0$ such that:
$$\left\|u\right\|_{W^{2, p}_\delta} \leq C \left(\left\|P_\epsilon(u)\right\|_{L^p_\delta} + \left\|u\right\|_{L^p_{\delta'}}\right)\quad \forall u \in W^{2, p}_\delta.$$
Indeed, if $F \subset W^{2, p}_\delta$ is a complementary subspace to $\ker P_\epsilon$, the norm $\left\|P_\epsilon(.)\right\|_{L^p_\delta} + \left\|.\right\|_{L^p_{\delta'}}$ is comparable on $F$ with $\left\|.\right\|_{W^{2, p}_\delta}$ and, $\ker P_\epsilon$ being finite dimensional, the norm $\left\|.\right\|_{L^p_{\delta'}}$ is comparable to $\left\|.\right\|_{W^{2, p}_\delta}$ on $\ker P_\epsilon$. Finally, $P = P_\epsilon + n \left|L\right|^2 \left(1-\psi\left(\frac{\rho}{\epsilon}\right)\right)$ so
$$
\left\| P(u) \right\|_{L^p_\delta}
	\leq \left\| P_\epsilon(u) \right\|_{L^p_\delta} + \left\| n \left|L\right|^2 \left(1-\psi\left(\frac{\rho}{\epsilon}\right)\right) u \right\|_{L^p_\delta}
	\leq C \left(\left\| P_\epsilon(u) \right\|_{L^p_\delta} + \left\| u \right\|_{L^p_{\delta'}}\right),
$$
because $n \left|L\right|^2 \left(1-\psi\left(\frac{\rho}{\epsilon}\right)\right)$ has a compact supprt ($C$ is a constant that depends only on $M, g, p, \delta$ and $\delta'$ and can vary from line to line). We have proven:
$$\left\|u\right\|_{W^{2, p}_\delta} \leq C \left(\left\|P(u)\right\|_{L^p_\delta} + \left\|u\right\|_{L^p_{\delta'}}\right)\quad \forall u \in W^{2, p}_\delta.$$
A proof similar to the one done in \cite{Gicquaud} shows that the kernel of $P$ coincides with the kernel of $P : W^{2, 2}_0 \to L^2_0$ and that if this kernel is reduced to $\{0\}$, then $P : W^{2, p}_\delta \to L^p_\delta$ is an isomorphism. The Lemma is then proved for $m > 0$ by applying elliptic regularity in M\"obius charts (see e.g. \cite{LeeFredholm}).
\end{proof}

\newpage
We can now state the main result of this article:

\begin{theorem}[Linearization stability of the constraint operator]\label{thmLinStab} Let $(M, g, K)$ be a triple such that  $(M, g)$ is a $\cC^{l, \beta}$-asymptotically hyperbolic manifold with $l+\beta \geq 2$ and $K$ a symmetric 2-tensor, then, assuming that
\begin{enumerate}
\item $\tr_g K$ is constant on $M$,
\item $J_i = \nabla_i \left( \tr_g K \right) - \nabla^j K_{ij} = 0$,
\item $L \in \cC^{m, \alpha}_0$ with $\left|L\right|^2 \in \cC^{0, \alpha}_{\delta'}$ for some $\delta' > 0$,
\item the $L^2$ kernel of the operator $P : u \mapsto -(n-1) \Delta u - \scal~u + n \left|L\right|^2 u$ is trivial,
\end{enumerate}
then the constraint operator $\Phi_r$ is linearization stable at $(M, g, K)$ between the following spaces:
\begin{itemize}
\item if $0 \leq \delta < n$ and $m+2+\alpha \leq l + \beta$,
$$\cC^{m+2, \alpha}_\delta \times \cC^{m+1, \alpha}_\delta \to \cC^{m, \alpha}_\delta   \times \cC^{m, \alpha}_\delta,$$

\item if $\delta \geq 0$,$\left|\delta + \frac{n-1}{p} - \frac{n-1}{2} \right| < \frac{n+1}{2}$, $n < (m+2)p$ et $m+2\leq l $
$$W^{m+2, p}_{\delta} \times W^{m+1, p}_{\delta} \to W^{m, p}_{\delta} \times W^{m, p}_{\delta},$$

\item for any $\delta$ such that $\left|\delta + \frac{n-1}{p} - \frac{n-1}{2} \right| < \frac{n+1}{2}$, if $n < m p$, $m+2 \leq l + \beta$ and $\left\lfloor\frac{m}{2}\right\rfloor \leq k + \alpha \leq l + \beta$
$$\Wtil^{m+2, p; k+2, \alpha}_{\delta} \times \Wtil^{m+1, p; k+1, \alpha}_{\delta} \to \Wtil^{m, p; k, \alpha}_{\delta} \times \Wtil^{m, p; k, \alpha}_{\delta}.$$
\end{itemize}
\end{theorem}

\begin{proof}
The spaces appearing in the Theorem are such that the constraint operator $\Phi$ is differentiable at $(g, K)$. Thus, by Proposition \ref{propLinStab2}, it is enough to prove that the composition  $\varphi \circ f$ is an isomorphism. Under the assumption of the Theorem, the composition $\varphi \circ f$ is given by:
$$
f \circ \varphi (u, \xi) = \left(
\begin{array}{c}
- (n-1) \Delta u - \scal~u + n \left| L\right|^2_g u - 2 \left\langle L, \rlie_\xi g \right\rangle\\
 \\
\Delta_{TT} \xi_i
\end{array}
\right).
$$
This is a differential operator between the following spaces:
$$
\left\lbrace
\begin{array}{rll}
\cC^{m+2, \alpha}_\delta \times \cC^{m+2, \alpha}_\delta & \to \cC^{m, \alpha}_\delta \times \cC^{m, \alpha}_\delta &\text{if $\delta \in (-1, n)$}\\
W^{m+2, p}_{\delta}	\times W^{m+2, p}_{\delta}		 & \to W^{m, p}_{\delta} \times W^{m, p}_{\delta} &\text{if $\left|\delta + \frac{n-1}{p} - \frac{n-1}{2} \right| < \frac{n+1}{2}$,}
\end{array}
\right.
$$
if $L \in \cC^{m, \alpha}_0$. By solving first the second line of:
\begin{equation}\label{eqLinStab}
\varphi \circ f (u, \xi) = \left(
\begin{array}{c}
2 \delta \rho\\
 \\
-\delta J_i
\end{array}
\right)
\end{equation}

using Lemma \ref{lmIsoTT}, then the first line using Lemma \ref{lmIsoHam}, we prove that the operator $\varphi \circ f$ is an isomorphism is an isomorphism between the spaces mentionned earlier. This proves linearization stability for H\"older and Sobolev spaces. In order to treat the mixed case, remark that, if $K \in \cC^{m, \alpha}_0$, $\varphi$ is a continuous linear map between the following spaces:
$$f : \Wtil^{m+2, p; k+2, \alpha}_{\delta} \times \Wtil^{m+2, p; k+2, \alpha}_{\delta}
	\to \Wtil^{m+2, p; k+2, \alpha}_{\delta} \times \Wtil^{m+1, p; k+1, \alpha}_{\delta},$$
and $\varphi$ :
$$\varphi : \Wtil^{m+2, p; k+2, \alpha}_{\delta} \times \Wtil^{m+1, p; k+1, \alpha}_{\delta}
	\to \Wtil^{m, p; k, \alpha}_{\delta} \times \Wtil^{m, p; k, \alpha}_{\delta}.$$
As a consequence, the composition:
$$\varphi \circ f : \Wtil^{m+2, p; k+2, \alpha}_{\delta} \times \Wtil^{m+2, p; k+2, \alpha}_{\delta} \to \Wtil^{m, p; k, \alpha}_{\delta} \times \Wtil^{m, p; k, \alpha}_{\delta}$$
is an isomorphism. Indeed, it is obvious that $\varphi \circ f$ is injective. We prove that it is also surjective. Let $(2 \delta \rho, -\delta J_i) \in \Wtil^{m, p; k, \alpha}_{\delta} \times \Wtil^{m, p; k, \alpha}_{\delta}$. There exists solutions $(u_1, \xi_1) \in W^{m+2, p}_{\delta} \times W^{m+2, p}_{\delta}$ and $(u_2, \xi_2) \in \cC^{k+2, \alpha}_0 \times \cC^{k+2, \alpha}_0$ to equation \eqref{eqLinStab}. Choosing $\delta' \leq \delta$ such that $-1 < \delta' + \frac{n-1}{p} < 0$ close enough to $-1$, the solution to \eqref{eqLinStab} is unique in $W^{2, p}_{\delta'}$ and $W^{m+2, p}_\delta, \cC^{k+2, \alpha}_0 \subset W^{2, p}_{\delta'}$ so $(u_1, \xi_1) = (u_2, \xi_2) \in \Wtil^{m+2, p; k+2, \alpha}_{\delta} \times \Wtil^{m+2, p; k+2, \alpha}_{\delta}$. This proves linearization stability for the $\Wtil$ spaces.
\end{proof}

\subsection{Examples of applications of Theorem \ref{thmLinStab}}\label{secLinStabExemples}

In this section, we give two classes of spaces that are linearization stable.

\begin{prop}[Linearization stability of vacuum spaces]\label{propLinStabVide} Let $(M, g, K)$ be Cauchy data where $(M, g)$ is a $\cC^{l, \beta}$-asymptotically hyperbolic manifold and $K$ a symetric 2-tensor on $M$ such that:
\begin{enumerate}
\item $\tr_g K$ is constant on $M$,
\item $(M, g, K)$ satisfies the vaccum constraint equations: $\rho = 0$ and $J_i = 0$, with $\left(1-\frac{1}{n}\right) \left(\tr_g K\right)^2 - 2 \Lambda_c = n (n-1)$,
\item $\left|L\right|^2 \in \cC^{0, \alpha}_{\delta'}$ for some $\delta' > 0$,
\end{enumerate}
then the constraint equations are linearization stable at $(M, g, K)$ in the sense of Theorem \ref{thmLinStab}.
\end{prop}

\begin{proof} It is enough to prove that the operator $P : u \mapsto - (n-1) \Delta u - \scal~u + n \left| L\right|^2_g u $ has trivial $L^2$-kernel. The vacuum Hamiltonian constraint reads
$$\scal - 2 \Lambda_c + \left(\tr_g K\right)^2 - \left|K\right|^2_g = 0$$
so
$$\scal = 2 \Lambda_c - \left(\tr_g K\right)^2 + \left|K\right|^2_g =  2 \Lambda_c - \left(1 - \frac{1}{n}\right) \left(\tr_g K\right)^2 + \left|L\right|^2_g$$
$$-\scal + n \left|L\right|^2_g = n(n-1) + (n-1) \left|L\right|^2_g \geq n(n-1).$$

If $u$ is an element of the $L^2$-kernel of $P$, this implies :

$$0 = \int_M u \left( P u \right) \geq \int_M \left(\left|\nabla u\right|^2_g + n(n-1) u^2 \right).$$

So this proves that $$ \int_M u^2 = 0.$$ Finally $u = 0$.
\end{proof}

Remark that this case encompasses the hyperbolic sections of the Minkowski space-time ($\Lambda_c = 0$ and $\tr_g K = \pm 1$) and natural sections of the anti-de Sitter space-time (see e.g. \cite{Hawking} ou \cite{WolfConstantCurvature} for the definition of this space-time). However, the anti-de Sitter space-time is not globally hyperbolic so, in the case $\Lambda_c < 0$, we only obtain the linearization stability of the Cauchy development of $M$. The second application of Theorem \ref{thmLinStab} extends the corresponding result in \cite[proof of Theorem 1]{BrunaGirbau3} :

\begin{prop}[Linearization stability of the Friedmann--Lema\^itre--Robertson-Walker space-times]\label{propLinStabFLRW} Let $(M, g, K)$ be a $\cC^{l, \beta}$-asymptotically hyperbolic manifold such that:
\begin{enumerate}
\item The energy density $\rho$ is constant on $M$: $\rho = \rho_0$ and $J_i = 0$,
\item $L = \mathring{K} = 0$,
\item $\scal = -n(n-1)$.
\end{enumerate}
then the constraint equations are linearization stable at $(M, g, K)$ in the sense of Theorem \ref{thmLinStab}.
\end{prop}

This Proposition includes natural spatial sections of the Friedmann--Lema\^itre--Robertson--Walker spaces with $K = -1$ (i.e. whose induced metric corresponds up to rescaling to the hyperbolic metric). The condition that $\tr_g K$ is constant does not appears in the proposition being a direct consequence of $L = 0$ and $J_i = 0$ by the momentum constraint. The proof is similar to the proof of the previous proposition.

\subsection{Counterexamples}\label{secLinStabCtrEx}
In this section, we give examples proving that the constraint equations are not linearization stable outside the weight interval. We begin by giving a construction of compactly supported TT-tensors on $\bR^n$:

\begin{prop}\label{propTTcompact}
Let $\Omega \subset \bR^n$ be a non-empty relatively compact open subset. There exists a non trivial compactly supported TT-tensor on $\Omega$.
\end{prop}

If $\Omega = \bB^n$ the open unit ball of $\bR^n$, using the conformal transformation formula for TT-tensors (see e.g. \cite[Proposition 4.1]{Gicquaud}), one can construct non-zero compactly supported TT-tensors on the hyperbolic space $\bH^n$. Before giving the proof of this Proposition, we give two examples of instability of the constraint equations outside the weight interval:

\begin{prop}[Instability of the constraint equations outside the regularity interval on the anti-de Sitter space]\label{propInstabAdS} Let $\delta > n$. The constraint map $\Phi_r$ is not linearization stable at $(\bH^n, b, K=0)$, where $b$ is the hyperbolic metric, for $\Lambda_c =- \frac{n(n-1)}{2}$, when seen as a map between the following spaces: $$\cC^{2, 0}_\delta \times \cC^{1, 0}_\delta \to \cC^{0, 0}_\delta \times \cC^{0, 0}_\delta.$$
\end{prop}

\begin{proof}
Let $h$ be a non-zero compactly supported TT-tensor on $\bH^n$. It is easy to see that $(h, k=0)$ is a solution of the linearized constraint equations $\varphi(h, k) = (0, 0)$ for $K=0$, $\rho = 0$, $J = 0$ and $\Lambda_c = - \frac{n(n-1)}{2}$. Assume that there exists one-parameter families $g(\lambda)$ and $K(\lambda)$ satisfying the vacuum constraint equations such that $g(\lambda) - b \in \cC^{2, \alpha}_\delta$ (where $b$ is the hyperbolic metric), $K(\lambda) - 0 \in \cC^{1, \alpha}_\delta$ with $g(\lambda) = b + \lambda k + o(\lambda)$ and $K(\lambda) = o(\lambda)$. If $g(\lambda)$ is a one-parameter family of metrics such that $g(0) = b$ and $\diff{g}{\lambda}(0) = h$ then $\diff{}{\lambda}\left(\ric_{g(\lambda)} + (n-1) g(\lambda) \right)(\lambda = 0) \neq 0$. Indeed, if this variation was zero, the following elliptic equation for $h$ would hold: $$-\frac{1}{2} \Delta_L h + (n-1) h = 0,$$ where $\Delta_L$ is the Lichnerowicz Laplacian (see e.g. \cite{LeeFredholm}). However this equation cannot admit any compactly supported solution $h$ (this is a consequence of the estimate \cite[Lemma 7.13]{LeeFredholm}). It can also be easily seen that the energy and the momentum of $(g(\lambda), K(\lambda))$ are zero so using \cite[Theorem 1.4]{MaertenPositiveEnergy}, the $(g(\lambda), K(\lambda))$ are hypersurfaces in the anti-de Sitter space. In particular, the Gauss equation traced on spatial coordinates leads to $$\ric_{g(\lambda)} + (n-1) g(\lambda) = \left(\tr_{g(\lambda)} K(\lambda)\right) K(\lambda) - K^2(\lambda) = o(\lambda^2),$$ this contradicts the fact that $\diff{}{\lambda}\left(\ric_{g(\lambda)} + (n-1) g(\lambda) \right)(\lambda = 0) \neq 0$.
\end{proof}

The same construction together with the positive energy theorem for asymptotically Euclidean manifolds (see e.g. \cite{ChruscielMaerten} and references therein) provides the following counterexample for asymptotically flat Cauchy surfaces:

\begin{prop}[Instability of the constraint equations on $\bR^n$]\label{propInstabAE}
Let $\delta > 0$, the constraint operator $\Phi_r$ is not linearization stable at $(\bR^n, e, K = 0)$, where $e$ denotes the Euclidean metric on $\bR^n$, between the following spaces:
$$\cC^2_{n-2+\delta} \times \cC^1_{n-1+\delta} \to \cC^0_{n+\delta} \times \cC^0_{n+\delta},$$
where the spaces $\cC^k_\alpha$ are defined (for symmetric 2-tensors) by $$\cC^k_\alpha = \left\{T \in \cC^k_{loc} \vert T_{ij} = O(r^{-\alpha}), \partial_{l_1} T_{ij} = O(r^{-\alpha-1}), \ldots, \partial_{l_1} \cdots \partial_{l_k} T_{ij} = O(r^{-\alpha-k}), \forall~i, j, l_1, \ldots, l_k \in \{1, 2, \ldots, n\} \right\},$$ $r$ being the Euclidean distance from the origin.
\end{prop}

\begin{proof}[Proof of Proposition \ref{propTTcompact}] Let $T_0$ be a compactly supported traceless symmetric 2-tensor. Define
$$
\left\lbrace
\begin{aligned}
\alpha & = \frac{n}{n-1} \partial^s \partial^t T_{0 st}\\
\psi_j & = \Delta \partial^i T_{0 ij} - \left( 1 - \frac{1}{n} \right) \partial_j \alpha,
\end{aligned}
\right.
$$
then it is easily verified that the tensor $T_{ij}$ defined as $$T_{ij} = \Delta \left( \Delta T_{0 ij} \right) - \left(\partial_i \partial_j \alpha - \frac{1}{n} \Delta \alpha \delta_{ij} \right) - \left(\partial_i \psi_j + \partial_j \psi_i \right)$$ is a TT-tensor. We give now an example of non-trivial compactly supported TT-tensor obtained using this construction. Choose a symmetric matrix $M$ with zeros on the diagonal and $\chi$ smooth compactly supported function whose support is contained in $\Omega$ and such that $\chi = 1$ on a compact $K$ with non-empty interior. Define $T_0= \chi \sum_k (x_k)^4 M$. A simple calculation shows that on $K$, $$T_{ij} = 24 (n-2) M_{ij} \neq 0.$$
\end{proof}

\rqs
\begin{enumerate}
\item This construction looks rather non natural. The underlying idea is to pass to the Fourier transform. Then subtract to $\widehat{T_0}$ an element of the form $\xi_i \hat{\psi}_j + \xi_j \hat{\psi}_i -\frac{2}{n} (\xi \cdot \hat{\psi}) \delta_{ij}$ (which is the Fourier transform of $\rlie_{\psi^\sharp} \delta$). And finally clear the denominators which are of the form $|\xi|^2$ or $|\xi|^4$ by multiplying by $|\xi|^4$.
\item There exists other constructions of compactly supported TT-tensors in dimension 3: \cite{CorvinoPenrose} (based on Hodge duality) and \cite{DainFriedrich} (using spherical harmonics). These constructions can lead to a parametrization of the set of compactly supported TT-tensors.
\end{enumerate}

\bibliographystyle{amsalpha}
\bibliography{biblio}

\end{document}